\documentclass[aps,showpacs]{revtex4}
\usepackage{graphicx}
\begin{document}
\title{Strength functions, entropies and duality in weakly to strongly
interacting fermionic systems}
\author{D. Angom}
\affiliation{Physical Research Laboratory, Ahmedabad \,\,380 009, India }
\author{S. Ghosh}
\affiliation{Physical Research Laboratory, Ahmedabad \,\,380 009, India }
\author{V.K.B. Kota \footnote{
corresponding author. Fax: 91-79-6301502. {\it e-mail address}: 
vkbkota@prl.ernet.in (V.K.B. Kota)}}
\affiliation{Physical Research Laboratory, Ahmedabad \,\,380 009, India }

\begin{abstract}

We revisit statistical wavefunction properties of finite systems of 
interacting fermions in the light of strength functions and their
participation  ratio and information entropy. For weakly interacting fermions
in a mean-field with random two-body  interactions of increasing strength
$\lambda$,  the strength functions $F_k(E)$ are well known to change, in the
regime where level fluctuations follow Wigner's surmise, from Breit-Wigner to
Gaussian  form. We propose an ansatz for the function describing this
transition which we use to investigate the participation ratio  $\xi_2$ and
the information entropy $S^{\rm info}$ during  this crossover, thereby
extending the known behavior valid in the Gaussian domain into much of the
Breit-Wigner domain. Our method also allows us to derive the scaling law for
the duality point $\lambda = \lambda_d$, where $F_k(E)$, $\xi_2$ and $S^{\rm
info}$ in both  the weak ($\lambda=0$) and strong mixing ($\lambda = \infty$)
basis  coincide as $\lambda_d \sim 1/\sqrt{m}$, where $m$ is the number of
fermions. As an application, the ansatz function for strength functions is
used in describing the Breit-Wigner to Gaussian transition seen in neutral
atoms CeI to SmI with valence electrons changing from 4 to 8. \\


\end{abstract}


\maketitle{}

\def\pr{\prime }
\def\be{\begin{equation}}
\def\lan{\left\langle}
\def\ran{\right\rangle}
\def\ee{\end{equation}}
\def\barr{\begin{array}}
\def\earr{\end{array}}
\def\non{\nonumber}
\def\nn8{\nonumber\\[15pt]}
\def\l{\left}
\def\ri{\right}
\def\dis{\displaystyle}
\def\ed{\end{document}}
\def\cg{\cal{G}}
\def\ce{\cal{E}}
\def\la{\mathrel{\mathchoice {\vcenter{\offinterlineskip\halign{\hfil
$\displaystyle##$\hfil\cr<\cr\sim\cr}}}
{\vcenter{\offinterlineskip\halign{
\hfil$\textstyle##$\hfil\cr<\cr\sim\cr}}}
{\vcenter{\offinterlineskip\halign{
\hfil$\scriptstyle##$\hfil\cr<\cr\sim\cr}}}
{\vcenter{\offinterlineskip\halign{
\hfil$\scriptscriptstyle##$\hfil\cr<\cr\sim
\cr}}}}}
\def\ga{\mathrel{\mathchoice {\vcenter{\offinterlineskip\halign{\hfil
$\displaystyle##$\hfil\cr>\cr\sim\cr}}}
{\vcenter{\offinterlineskip\halign{
\hfil$\textstyle##$\hfil\cr>\cr\sim\cr}}}
{\vcenter{\offinterlineskip\halign{
\hfil$\scriptstyle##$\hfil\cr>\cr\sim\cr}}}
{\vcenter{\offinterlineskip\halign{
\hfil$\scriptscriptstyle##$\hfil\cr>\cr\sim
\cr}}}}}
\oddsidemargin 0.0in \evensidemargin 0.5in
\marginparwidth 40pt \marginparsep 10pt
\topmargin 0in \headsep .5in
\textheight 8.6in \textwidth 6in
\brokenpenalty=10000
\parindent 0.2in

\section{Introduction}

There are many physical systems which are statistically well described by 
so-called embedded ensembles of  fermions, representing particles  subjected
to a one-body mean-field potential (defining a set of single-particle
levels), and interacting with a random, two-body potential. Examples include
heavy nuclei \cite{Ko-01,Bro-81,Ks-01,Ze-02}, natural \cite{Fl-99,An-03}, or
artificial atoms (quantum dots) \cite{Ja-01,Al-00}, and nanometer-scale
metallic grains \cite{Pa-02}. Similar situations of randomly interacting 
spin systems occur in the study of spin-glass systems \cite{Mez-87}, and in
the context of quantum information and quantum computation \cite{Qc-00}. 
These embedded ensemble are defined as ensembles of Hamiltonians $\{H\} =
h(1) + \lambda \{V(2)\}$, where $\{\ldots\}$ denotes an ensemble
distribution, $h(1)=\sum_i \epsilon_i n_i$ is a fixed one-body operator  (one
can also consider an ensemble $\{h(1)\}$ defined by a  probability
distribution $P(\epsilon_i)$) defined by the single-particle energies
$\epsilon_i$ with average spacing $\Delta$ which sets the energy scale (one
can thus set $\Delta=1$ without loss of generality), and $n_i$ is the number
operator for  the single-particle state $\l. \l| i\ri. \ran$. Similarly
$V(2)$ is the random two-body interaction with its two-particle matrix
elements chosen as independent Gaussian variables with zero center and unit
variance. Thus, for $m$ fermions in $N$ single-particle  states, $\{H\}$ is a
one plus two-body random matrix ensemble (called embedded ensemble of
(1+2)-body interactions[EGOE(1+2)]) \cite{Ko-01,Bro-81} defined by the
parameters $(m,N,\lambda)$,  where  $\lambda$ is the interaction strength.
For convenience, we only consider here EGOE(1+2) for spinless fermions, but
extensions to particles with intrinsic angular momentum have also been 
considered \cite{Ja-01,Pa-02,Kok-02}. In such a case, the size of Hilbert
space is given by $d=\left(^N_m\right)$ and another important parameter is the
connectivity $K$ giving the number of directly coupled states as $K=1+m
(N-m)+m (m-1) (N-m) (N-m-1)/4$.

Because of its broad relevance to many, a priori different, finite quantum
systems,  EGOE(1+2)'s have been investigated in detail by many research
groups in the recent past  
\cite{Ko-01,Bro-81,Ks-01,Ja-01,Al-00,Pa-02,Qc-00,Kok-02,Fl-96,Fl-97,Ab-90,
Ja-97,Gs-97,Fl-00,Ja-02,Ks-02}. Most investigations used analytical methods
extrapolating from the weak and the strong coupling limit, and relied on
numerical calculations in the regime of intermediate values of $\lambda$.
Focusing on the statistical spectral and wavefunction properties, the
dominant features of EGOE(1+2) that emerged from those investigations can be
summarized as:

\begin{enumerate}

\item There is a marker $\lambda_c$, such that for $\lambda > \lambda_c$ the
many-body level spacing  distribution becomes close to that of the Gaussian
Orthogonal Ensemble (GOE) of random matrices \cite{Mehta}, while for $\lambda
< \lambda_c$ the level fluctuations are close to Poisson.  It was further
established that $\lambda_c \propto 1/m^2N$ \cite{Ab-90,Ja-97};  specifically
for $m=6$ and $N=12$, $\lambda_c \simeq 0.06$ \cite{Ko-01}.

\item As $\lambda$ increases from $\lambda=0$, the strength function (to be
defined below in Section 2) undergoes a crossover from a delta-peak, first to
a Breit-Wigner (BW), then to a Gaussian form. Related to that  crossover,
there are two markers $\lambda_{F}^{(1,2)}$ such that, for $\lambda_F^{(1)}
\leq  \lambda \leq \lambda_F^{(2)}$,  the strength function is well
approximated by a BW form  \cite{Ko-01,Ks-01,Fl-96,Fl-97,Gs-97,Fl-00,Ja-02};
the BW form emerges above $\lambda_F^{(1)}$, which is exponentially smaller
in $m$ and $N$ than $\lambda_c$. In particular, the BW form occurs while the
spectral fluctuations are still Poissonian. The   $\lambda >
\lambda_{F}^{(2)}$ region (with full GOE spectral and wavefunction
properties) is called the Gaussian  domain \cite{Ks-01}. From now on we put
$\lambda_F=\lambda_F^{(2)}$; note that $\lambda_F >> \lambda_c$. Arguments
based on BW spreading widths give  $\lambda_F \propto 1/\sqrt{m}$
\cite{Fl-97,Ja-02}; for the specific case  $m=6$ and $N=12$, $\lambda_F 
\simeq 0.2$ \cite{Ks-01}.

\item In the Gaussian domain, the participation ratio (PR) $\xi_2$ and  the
exponential of the information entropy ($\exp[S^{\rm info}(E)]$) (both
quantities will be defined in Section 3 below) take Gaussian forms when
plotted as a function of the energy \cite{Ks-01}. The variances of these
Gaussian  are $(1+\zeta^2)/(2\zeta^2)$ and $1/\zeta^2$  respectively, where
$\zeta^2 = \sigma^2_h(m)/[\sigma^2_h(m) + \lambda^2 \sigma^2_V(m)]$;
$\sigma_h(m)$ is the spectrum width produced by $h(1)$ in the total
$m$-particle space and similarly $\sigma_V(m)$ is the width produced by
$V(2)$. Also, in the BW region,the  PR is given by the ratio of the
spreading width and the spacing between directly (by $V(2)$) connected states
and $S^{\rm info} \sim  \ln ({\mbox{PR}})$ \cite{Gs-97} (see also
\cite{Ja-95}).

\item There is a fourth marker $\lambda_d$ such that at $\lambda=\lambda_d$
the strength functions, PR and $S^{\rm info}$ expressed in either  the
$h(1)$ (i.e. $\lambda=0$) and $V(2)$ (i.e. $\lambda=\infty$)  basis will
coincide. This is accompanied by a duality transformation relating the values
of those quantities in the  $h(1)$ basis to those in the $V(2)$ basis by 
$\lambda \rightarrow \lambda_d^2/\lambda$ \cite{Ja-02}. In Section 5 the 
$(m,N)$ dependence of $\lambda_d$ will be shown to  be $\lambda_d \sim
1/\sqrt{m}$ (correcting the previously postulated result $\lambda_d \sim
1/m^{1/4}$ \cite{Ja-02}); for $m=6$ and $N=12$, as we see ahead, $\lambda_d
\simeq 0.3$.

\end{enumerate}

\noindent In addition,  another important result for EGOE(1+2) is that the
smoothed (ensemble averaged) density of states takes on a Gaussian form 
independently of $\lambda$ \cite{Ko-01,Bro-81,Be-01}. Our purpose in this
paper is to bring completion to the investigations related to the points
(1)-(4) above. In particular, we will introduce an interpolating function
for strength functions for the BW to Gaussian  transition and apply it to
neutral CeI to SmI atoms. Now we will give a preview. 

In Section 2 we discuss a variant of the well known Student's 
t-distribution \cite{Ke-69} (hereafter called $F_{k:BW-\cg}(E)$) and show
that it is well suited for describing the BW to Gaussian transition, as a
function of one parameter $\alpha$. Numerical calculations allow to
establish a one-to-one correspondence  between $\alpha$ and the interaction
strength $\lambda$. In Section 3, the resulting $F_{k:BW-\cg}(E)$ is used to
calculate both PR and $S^{\rm info}$, and comparison is made with direct
numerical calculations of these quantities as a function of $\lambda$, over
the full range of variation of $\lambda$,  thereby extending previous
similar investigations which were restricted to either the Gaussian 
\cite{Ks-01} or BW \cite{Gs-97,Ja-02} domains.  Additional structures in the
wave functions can be captured by the structural entropy $S_{\rm str} \equiv
S^{\rm info} - \ln \xi_2$, which measures the amount of information
contained in the tails of the strength functions.  Results of an analysis of
$S_{\rm str}$ is also given in Section 3. In Section 4, the
$F_{k:BW-\cg}(E)$ is applied in the analysis of the BW to Gaussian
transition one observes as we go from neutral CeI atom to SmI atom.   In
Section 5, the existence of a duality transformation in EGOE(1+2) (which was
established in Ref. \cite{Ja-02}) is discussed and it is shown that, using
the results for PR and $S^{\rm info}$ in the Gaussian  domain, the duality
point $\lambda_d \sim 1/\sqrt{m}$. Conclusions and final comments are given
in Section 6.

\section{Interpolating function for BW to Gaussian transition in strength
functions}

Given the mean-field $h(1)$ basis states $\l.\l|k\ri.\ran$ and the expansion 
of the eigenstates $\l.\l|E\ri.\ran$ as $\l.\l|E\ri.\ran =\sum_k\,C^E_k\,
\l.\l|k\ri.\ran$, the strength function $F_k(E)$ is defined by
\be
F_k(E) = \dis\sum_{E^\prime}\,\l| C^{E^\prime}_k\ri|^2 \delta(E-E^\prime) =
\langle \l| C^E_k\ri|^2 \rangle \,\l(d\,\rho^H(E)\ri)
\ee
In Eq. (1), $\langle \ldots \rangle$ indicates an ensemble average,
$d=\left(^N_m\right)$ is the $m$-particle space dimension and $\rho^H(E)$  is
the normalized (and ensemble averaged) density of states. As mentioned in 
Section 1, $\rho^H(E)$ is in general a Gaussian (often the superscript  $H$
is dropped), 
\be
\rho^H(E) = \dis\frac{1}{\dis\sqrt{2\pi}\,\sigma_H(m)}
\,\exp-\frac{{\hat{E}}^2}{2}\;;\;\;\hat{E}=(E-\epsilon_H(m))/
\sigma_H(m)
\ee
where $\epsilon_H(m)=\lan H \ran^m$ is the spectrum centroid and similarly
$\sigma_H(m)$ is the spectral width. The BW and Gaussian (denoted by $\cg$)
forms of $F_k(E)$ are,
\be
F_{k:BW}(E) = \dis\frac{1}{2 \pi}\, \dis\frac{\Gamma_k}{(E-E_k)^2 +
\Gamma_k^2/4},\;\;\;F_{k:\cg}(E) = \dis\frac{1}{\dis\sqrt{2
\pi}\,\sigma_k} \,\exp-\frac{(E-E_k)^2}{2 \sigma_k^2}
\ee
where $E_k = \lan k | H | k \ran$. With $p=\int^{{\ce}_p}_{-\infty}\,F_k(E)
dE$, the spreading width $\Gamma_k={\ce}_{3/4} - {\ce}_{1/4}$. 
Similarly the variance
of $F_k$ is $\sigma_k^2=\lan k \mid H^2 \mid k \ran - (\lan k \mid H \mid k 
\ran)^2$. Expressions for $\Gamma_k$ and $\sigma_k$ in terms of $(m, N,
\lambda)$ are given in Sections 5 and 3 respectively. Both spreading width
$\Gamma_k$ of the BW and $\sigma_k$ of the Gaussian strength functions are
essentially independent of $k$ (provided one considers energies not too far
away from the center of the density of states)  \cite{Ko-01,Ks-01,Fl-96}.
Simultaneously, the energies $E$'s (of $H$) and $E_k$'s will have the same
centroid. Moreover, just as the state density $\rho^H(E)$, the $E_k$'s
density (denoted by $\rho^{H_k}(E_k)$) is also a Gaussian. These results are
used throughout this paper and without loss of generality the centroids of
$E$'s and $E_k$'s are set equal to zero. As it is discussed in detail in
\cite{Ks-01}, the width of  $\rho^{H_k}(E_k)$ is essentially due to $h(1)$
(with a small correction from $V(2)$ as explained in the Appendix in
\cite{Ks-01}) and $\sigma_k$ is generated by $V(2)$. Before proceeding
further it should be mentioned that the strength functions are basis
dependent and one can define strength functions in the $V(2)$ basis also. We
will turn to this question when discussing duality transformation  in Section
5.

For the BW to Gaussian transition we make the following ansatz for $F_k(E)$
\be
F_{k:BW-\cg}(E:\alpha,\beta)\,dE =
\dis\frac{(\alpha\beta)^{\alpha-\frac{1}{2}}\;\Gamma(\alpha)}{\dis\sqrt{
\pi}\;\Gamma(\alpha-\frac{1}{2})}
\;\dis\frac{dE}{\l((E-E_k)^2+\alpha \beta\ri)^\alpha}\;,\;\;\;\alpha \ge 1
\ee
The $F_{k:BW-\cg}$ in (4) gives BW for $\alpha=1$ and Gaussian  for $\alpha
\rightarrow \infty$ (this can easily be checked using Stirling's
approximation).  As required, it is normalized to unity for any positive
value of the continuous parameter $\alpha$. For $2\alpha-1$ a integer,
$F_{k:BW-\cg}$   gives the so called {\it Student's} $t$-distribution
\cite{Ke-69}, which is  well known in statistics. In particular, the {\it
Student's} distribution $f(x)$ with  a parameter $\nu$ given in  Table 5.7 of
\cite{Ke-69} reduces to (4)  with the change $\alpha=(\nu+1)/2$, $\nu$ a
positive integer, and $x \rightarrow \sqrt{\frac{2\nu}{ \nu+1}}\,(E - E_k)/
\sqrt{\beta}$. Note that the construction of $F_{k:BW-\cg}$ in Eq. (4) is
similar in spirit to the Brody distribution, interpolating between the
Poisson and Wigner-Dyson distributions for nearest neighbor spacing
distribution (NNSD) \cite{Bro-81}. Also, just as some groups use for the NNSD
a linear combination of Poisson and Wigner forms multiplied by $x$ and
$(1-x)$ respectively with $x$ being the mixing parameter, it is possible to
use $\mu F_{k:BW}(E) + (1-\mu) F_{k:\cg}(E)$ for the BW to Gaussian
transition with $\mu$ ($0 \leq \mu \leq 1$) being the  mixing parameter. This
simple form is not explored in this paper as it is  unlikely that a theory
for strength functions for EGOE(1+2) will give this  form. 

In $F_{k:BW-\cg}(E: \alpha,\beta)$, the parameter $\alpha$ is sensitive to 
shape changes, while the parameter $\beta$ supplies the energy scale over 
which $F_{k:BW-\cg}(E: \alpha,\beta)$ extends. Since we focus on the shape
transformation, $\alpha$ is the significant parameter.
First, it is easy to see that $F_{k:BW-\cg}(E: \alpha,\beta)$ is an even
function of $E-E_k$, so that all of its finite odd cumulants 
vanish (strictly speaking, the centroid is $E_k$ only for $\alpha > 1$; see
\cite{Ke-69}).  The variance of $F_{k:BW-\cg}$, defined only 
for $\alpha > 3/2$, is 
\be
\sigma^2(F_{k:BW-\cg}) = \l(\dis\frac{\alpha}{2 \alpha -3}\ri)\,\beta \;,
\;\;\;\; \alpha > 3/2.
\ee
and it is useful to recall that $\sigma^2(F_k) = \lambda^2 \sigma_V^2$ 
\cite{Ks-01}. For $\alpha > 3/2$ one can use (5) to fix $\beta$ while for
$\alpha \leq 3/2$, it is the spreading width $\Gamma$ (this is well defined
for all $\alpha$ values) that is useful for fixing the $\beta$ value. There
is no simple expression for $\Gamma$ as a function of $\alpha$ and $\beta$
but using (4) this can be calculated numerically. Just as Eq. (5), the excess
parameter (also known as Kurtosis) of $F_{k:BW-\cg}$ is
$\gamma_2=6/(2\alpha-5)$ for $\alpha > 5/2$. However this expression is not
useful in practice as the spectrum is always of finite range and this causes 
large deviations for $\alpha \approx 2-8$. Therefore it is more useful to
consider $\gamma_2$ with the spectrum ranging say  from $-a$ to $+a$. Then
Eq. (4), with proper normalization, gives,
\be
\gamma_2(a:\alpha,\beta)=\dis\frac{9}{5} \;\dis\frac{
_2F_1(\frac{5}{2},\alpha; \frac{7}{2};
-\eta^2)\; _2F_1(\frac{1}{2},\alpha; \frac{3}{2};
-\eta^2)}{\l[_2F_1(\frac{3}{2},\alpha; \frac{5}{2};-\eta^2)\ri]^2} -3
\ee
where $\eta^2=(a^2/\alpha\beta)$, and $_2F_1$ a hypergeometric function.

In Fig. 1 the results of EGOE(1+2) for $F_k(E)$ are compared, for the  $m=6$,
$N=12$ system at $E_k=0$ (i.e. in the middle of the band),  with the best fit
$F_{k:BW-\cg}$ for various values of $\lambda$.  In the fits, the $\beta$
values are fixed using Eq.(5) for $\lambda \ge 0.1$ (for these, $\alpha >
1.6$) and the spreading width $\Gamma$ is used ($\beta \sim \Gamma^2/4$) for
$\lambda=0.06$ (here $\alpha=1.2$). In the fits, also imposed is the
condition that the value of $\gamma_2$ calculated from Eq. (6) over the
spectrum   range should be close to the numerical EGOE(1+2) values. As it is
seen from Fig. 1, the fits are excellent over a wide range of $\lambda$
values; in the calculations only $\lambda \geq 0.06$ are considered (for the
system considered in Fig. 1, $\lambda_c \sim 0.06$). The variation of
$\alpha$ with $\lambda$ is shown in Fig. 2. The parameter $\alpha$ raises
slowly upto  $\lambda_F$ (note that $\lambda_F \sim 0.2$ for the EGOE(1+2)
system used in Fig. 1 \cite{Ks-01}) and then it starts rising sharply with
$\lambda$. Finally the $\alpha$ values start saturating after $\lambda >
\lambda_0=0.3$ (the saturation is artificial as determination of $\alpha$ for
$\lambda >> \lambda_0$ is difficult as here $F_k(E)$ will be very close to
Gaussian). The criteria $\alpha \sim 4$ and $\gamma_2 \sim 1$ appear to
define $\lambda_F$. Fig. 2 shows  that the BW to Gaussian transition is a
sharp transition and therefore studies in BW and Gaussian regimes can be
carried out independently, to a good approximation, 
as it is done in many papers before.  Now we will
apply $F_{k:BW-\cg}$ to study PR and $S^{info}$ in the region intermediate to
BW and Gaussian forms.

\section{Participation ratio and Information entropy  in the BW to Gaussian
transition region}

Two important measures of the complexity of eigenstates of 
interacting systems are the participation ratio and information entropy. 
As in the previous section, we expand the Hamiltonian eigenstates in the
noninteracting mean-field basis as $\l.\l|E\ri.\ran =\sum_k\,C^E_k\,
\l.\l|k\ri.\ran$. Then PR and information entropy are defined as
\be
\barr{rcl}
\xi_2(E) & = & \l\{\dis\sum_k\,|C_k^E|^4\ri\}^{-1}\;, \\
S^{\rm info}(E) & = & -\dis\sum_k\,\l|C^E_k\ri|^2 \ln \l|C^E_k\ri|^2\;. 
\earr
\ee
The subscript `2' denotes that $\xi_2$ is the second R\'{e}nyi  entropy
\cite{Im-02}. Qualitatively, $\xi_2$ counts the number  of
$\{|k\rangle\}$-basis components necessary to construct one  typical $|E
\rangle$-state, and is thus often referred to as the Number of Principal
Components (NPC). Obviously, both $\xi_2$ and $S^{\rm info}$ are basis
dependent, and could as well be  defined starting from another expansion.
Eq. (7) gives their  expression with respect to the $h(1)$ basis (i.e. the
$\l.\l| k \ri. \ran$'s in Eq. (7), as in Section 2, are $h(1)$ basis
states), consequently, $\xi_2$  and $S^{\rm info}$ give measures of the
spreading of eigenstates over the noninteracting basis as the many-body
interaction is made stronger and stronger. Below, in Section 5, 
we will deal with these
measures defined with respect to the $V(2)$  basis. Following Ref.
\cite{Ks-01} one can write $\xi_2$ and $S^{\rm info}$  in terms of the
strength functions,
\be
\barr{rcl}
\l\{\xi_2(E)/\xi_2^{GOE}\ri\}^{-1} & = & \dis\frac{1}{\l[\rho^H(E)\ri]^2}\;
\dis\int_{-\infty}^{\infty}\,dE_k\;\rho^{H_k}(E_k) \l[F_k(E)\ri]^2\;, \nn8
S^{\rm info}(E) - S^{\rm info}_{GOE} & = &
-\dis\frac{1}{\rho^H(E)}\;\dis\int_{-\infty}^{\infty}\,dE_k\;\rho^{H_k}(E_k)
F_k(E)\;\ln \dis\frac{F_k(E)}{\rho^H(E)}\;. 
\earr
\ee
An obvious notation refers to the GOE values, 
\be
\barr{rcl}
\xi_2^{GOE} & = & d/3\,, \\
\exp(S^{\rm info}_{GOE}) & = & 0.48 d \,.
\earr
\ee
In the Gaussian domain, with $F_k(E)$ being a Gaussian with centroid at
$E_k$ and width $\lambda \sigma_V(m)$, the integrals in Eq. (8) are easy to 
evaluate (note that $\rho^H(E)$ and $\rho^{H_k}(E_k)$ are zero centered
Gaussians with width $\sigma_H(m)$ and $\sigma_h(m)$ respectively) and they
give \cite{Ks-01},
\be
\barr{rcl}
\xi_2(E)/\xi_2^{GOE} & = & \dis\sqrt{1-\zeta^4}\;
\exp-\frac{\zeta^2 {\hat{E}}^2}{1+\zeta^2}\,, \nn8
\exp(S^{\rm info}(E) - S^{\rm info}_{GOE}) & = & \dis\sqrt{1-\zeta^2}\;
\exp\frac{\zeta^2}{2} \;\exp-\frac{\zeta^2 {\hat{E}}^2}{2}\,. 
\earr
\ee
The important parameter in Eq. (10) is the correlation coefficient $\zeta$.
For $\xi_2$  and $S^{\rm info}$ in $h(1)$ ($\lambda=0$) basis $\zeta \equiv
\zeta_0$ where
\be
\l(\zeta_0^{(m)}\ri)^2 = \dis\frac{\sigma^2_h(m)}{\sigma^2_h(m) + 
\lambda^2 \sigma^2_V(m)}
\ee
Strictly speaking $\sigma^2_h(m)$ in the numerator should be replaced by the
variance of $E_k$ energies and similarly the denominator by $\sigma^2_H(m)$.
Eq.(11) is obtained by recognizing that the former is very close to
$\sigma_h^2$ and the later is essentially $\sigma^2_h + \lambda^2
\sigma^2_V$.  In fact these results are valid in the dilute limit  ($m
\rightarrow \infty$, $N \rightarrow \infty$, $m/N \rightarrow 0$) and here
$h$ and $V$ are orthogonal. Even away from the dilute limit they remain to be
good approximations (see Fig. 3 ahead for a test). Propagation  formulas
\cite{Ko-01} for $\sigma^2_h(m)$ and $\sigma^2_V(m)$ are
\be
\barr{l}
\sigma_h^2(m) = \dis\frac{m(N-m)}{(N-1)}\;\sigma^2_h(1) = f^2 \Delta^2 \nn8
\lambda^2\,\sigma^2_V(m) =
\dis\frac{m(m-1)(N-m)(N-m-1)N(N-1)}{(N-2)(N-3)}\;\dis\frac{\lambda^2}{4} = 
g^2 \lambda^2
\earr
\ee
It is always possible to write $\sigma^2_h(1)$ in terms of $\Delta^2$ and 
for example for a uniform single particle spectrum,
\be
\sigma^2_h(1)=(N+1)(N-1)\; \dis\frac{\Delta^2}{12} 
\ee
As shown in Fig. 3 (for the $m=6,N=12$ system), results of the formulas 
(11,12) agree very well with numerical EGOE(1+2) values for $\zeta_0$. 

Substituting the interpolating $F_{k:BW-\cg}(E)$ for $F_k(E)$ in Eq. (8), 
one can study $\xi_2$ and $S^{\rm info}$ as a function of $\lambda$. The 
integral in Eq. (8) for $\xi_2$ can be simplified for $E=0$ and this gives
(for $\alpha > 3/2$),
\be
\barr{l}
\xi_2(E=0)/\xi_2^{GOE} = \nn8
 \l\{ \dis\sqrt{\dis\frac{2}{(2 \alpha -3)}}\;
\dis\frac{\Gamma^2(\alpha)}{\Gamma^2(\alpha-\frac{1}{2})}\;
\dis\frac{1}{\dis\sqrt{\zeta^2 (1-\zeta^2)}} \; U\l(\frac{1}{2},\; 
\frac{3}{2} - 2 \alpha,\; \frac{(2\alpha-3)(1-\zeta^2)}{2 \zeta^2}\ri)
\ri\}^{-1}
\earr
\ee
where $U(---)$ is hypergeometric-U function \cite{Ab-64}. The corresponding
result in the Gaussian domain (from Eq. (10)) is $\sqrt{1-\zeta^4}$. In
deriving Eq. (14), we used Eqs. (5) and (11) for  eliminating $\beta$ and
simplifying all the variances into $\zeta^2$. It is important to note that
Eq. (14) is valid only for $\alpha > 3/2$ as (5) is used in deriving this
formula. For $\alpha \leq 3/2$ a compact formula could not be derived but one
can use (8) for numerical evaluations. Similarly, in the case of $S^{\rm
info}(E=0)$ a simple formula like Eq. (14) could not  be obtained for any
$\alpha$ and once again here one can use (8) for numerical evaluations. For
$\xi_2(E=0)$ results from (14) for $\lambda \geq 0.08$ and  the result from
(8) for $\lambda=0.06$  are compared  with numerical EGOE(1+2) calculations
for the $m=6,N=12$ system in Fig. 4. In these calculations the
$\alpha$-values are read off from Fig. 2 and $\zeta^2$ from Fig. 3. Comparing
with the Gaussian domain results given by (10), it is seen that they are good
for $\lambda > \lambda_F$ as expected; these results again confirm that
$\lambda_F \sim 0.2$ for the $m=6,N=12$ system. The agreement between  (14,8)
and  the numerical calculations continue upto  $\lambda \sim 0.06$. For
$\lambda < \lambda_F$ as the BW structure is more dominant, there will be
more localization and hence $\xi_2$ decreases fast as $\lambda$ is decreasing
as seen in Fig. 4; finally they will approach zero for $\lambda \rightarrow
0$. The results based on (8) will not extend to the region $\lambda \la
\lambda_c$ as here the GOE  assumptions used in deriving these equations (see
\cite{Ks-01}) will fail. Finally, for $S^{\rm info}(E=0)$ the results are
similar to those shown in Fig. 4. This is not surprising as
in many numerical calculations (including the present calculations)  it is
seen that $S^{\rm info}(E) \sim \ln( \xi_2(E))$ and therefore only their
difference can capture the information not contained in the bulk of $S^{\rm
info}$ or PR. With this clue, recently it is argued \cite{Im-02}  that the
structural entropy $S_{\rm str}(E)=S^{\rm info}(E)-\ln(\xi_2(E))$ is an
important measure of complexity (in addition to $S^{\rm info}(E)$ or
$\xi_2(E)$) in eigenfunctions. More importantly $S_{\rm str}$ is free of
divergences associated with $S^{\rm info}$ and PR. Note that $\exp(S^{\rm
info}_{GOE})$ and $\xi_2^{GOE}$ are $0.48d$ and $d/3$ respectively and
therefore they diverge as the matrix dimension $d \rightarrow \infty$. For
interacting particle systems it is observed that $S_{\rm str}(E=0)$ vs
$\lambda$ (or the disorder in the Anderson  model\cite{Im-02}) exhibits a
peak. It is then of interest to examine $S_{\rm str}$ in terms of the
results given Section 2. First we consider $S_{\rm str}$ in the BW and
Gaussian domains.

For small $\lambda$, one can estimate $S_{\rm str}$ using the the BW
approach, i.e. taking $F_{k:BW}$ from Eq.(3), and  using $\lan |C_k^E|^2
\ran = F_{k:BW}(E) \Delta_m$, where $\Delta_m$ gives the many-body level
spacing. Inserting this into Eq.(7) and replacing the sums by integrals over
$E_k$ one then gets, by restricting to $E=0$, 
\be
S_{\rm str}(0) = \frac{1}{\pi} \int_{-\tan\,a}^{\tan\,a}
d \epsilon \frac{\ln (1+\epsilon^2)}{1+\epsilon^2} \;+\;
\ln \l[\frac{2a + \sin\,2a}{2 \pi}\ri]\; +\; 
\ln \l[\frac{\pi \Gamma}{2 \Delta_m}\ri] \l(\frac{2a}{\pi} -1\ri),\;\;
a=\arctan(2B/\Gamma)
\ee
where $B$ is the many-body bandwidth and $\Gamma \propto \lambda^2$ is the BW
width. The upper limit $S_{\rm str}(0) \rightarrow \ln 2 \sim 0.7$ follows 
by letting $B/\Gamma \rightarrow \infty$ (then $a \rightarrow \pi/2$) in
(15). With $\Gamma$ increasing with increasing $\lambda$,  the $S_{\rm
str}(0)$ starts decreasing from the maximum value. Similarly, in the 
Gaussian domain, using (10), one has
\be
S_{\rm str}(0) = \ln(1.44) + \dis\frac{1}{2}\l(\zeta^2 - \ln(1+\zeta^2)\ri)
\ee
It should be noted that $ S^{GOE}_{\rm str} \simeq \ln(1.44)$ independent of
$E$. An interesting observation (though its significance is not clear) is
that for $\lambda=0$ (then $\zeta=1$) the $S_{\rm str}$ is sum $S_{\rm str}$
for GOE and a Gaussian; note that for a Gaussian distribution, as shown in
\cite{Im-02}, $S_{\rm str}=\frac{1}{2}(1-\ln2)$. As $\lambda \rightarrow
\infty$ gives $\zeta=0$, $S_{\rm str}$ starts from GOE value (0.3689) for
very large $\lambda$ and then starts increasing with decreasing $\lambda$.

For the intermediate regime and for $\lambda \leq \lambda_c$ no analytical
results could be derived yet but numerical calculations give some insight.
Fig. 5 shows the EGOE(1+2) results for $S_{\rm str}(E=0)$ vs $\lambda$ in the
$h(1)$ basis and their comparison with the results from (8,14) where
$F_{k:BW-\cg}$ given by (4) is used.   The $S_{\rm str}$ is well described
for $\lambda >> \lambda_c$ and for the $m=6,N=12$ system considered in Fig. 5
the theory given by (8,14) is good upto $\lambda = 0.1$.  Comparing with Eqs.
(15,16), it is seen that the Gaussian domain result (16) describes the
results for $\lambda \geq \lambda_F$ while the BW result (15)  describes only
the trends for $\lambda$ between 0.1 and $\lambda_F$ (with the maximum
possible value for $S_{\rm str}$ being 0.7). More importantly, as seen from
Fig. 5   and also from Fig. 4 of \cite{Ja-02}, $S_{\rm str}$ exhibits a peak
around a $\lambda$ value not far from $\lambda_c$ marker and here the level
fluctuations will have Poisson component. Thus it is plausible that the peak
arises due to large spectral(and strength) fluctuations. A good theory for
$S_{\rm str}$ generating the observed peak is at present not available.

\section{BW to Gaussian transition in neutral atoms CeI to SmI} 

Realistic examples for the BW to Gaussian transition in strength functions
are expected to come from neutral lanthanide atoms. It is known from the
analysis  of CeI by Flambaum et al \cite{Fl-99} and PrI by Cummings et al 
\cite{cummings-3407-01} that the $F_k(E)$ for these atoms with 4 and 5
valance electrons respectively are close to BW while those of SmI with 8
valance electrons, as shown in \cite{An-03}, are close to Gaussian. This
series of atoms is completed by NdI and PmI with 6 and 7 valance electrons.
We made calculations for not only NdI and PmI but also for the other three by
using the same method and this is briefly discussed below before giving the
results. Also let us add that as $\lambda_F \propto 1/\sqrt{m}$, it is to be
expected that, with $m$ changing from 4-8, there should be BW to Gaussian
transition with $H$ fixed. This is indeed confirmed by the results discussed
ahead.

The ground state configurations of Ce, Pr, Nd, Pm  and Sm are $4f5d6s^2
(^1G_4)$, $4f^36s^2 (^4I_{9/2})$, $4f^46s^2 (^5I_{4})$,  $4f^56s^2
(^6H_{5/2})$ and $4f^66s^2 (^7F_{0})$ respectively. Coupling of the $5d$ and
$4f$ valence electrons produce several configurations and indicate  strong
configuration mixing. Previous work on Sm I \cite{angom-271-2001} and 
lanthanide series \cite{sekiya-012503-01} have established that an
appropriate  method of calculation is the multiconfiguration Dirac-Fock
(MCDF)  \cite{grant-23-87}, where an atomic state function (ASF) $|\overline{
\Gamma} PJM\rangle$  is approximated as a linear combination of configuration
state  functions ( CSFs) $|\gamma PJM\rangle$. That is  $|\overline{\Gamma}
PJM\rangle = \sum_k c_k|\gamma_r PJM\rangle$, where $c_k$s are the  mixing
coefficients, $P$, $J$ and $M$ are parity, total angular momentum and 
magnetic quantum numbers respectively, and $\overline{\Gamma}$ and $\gamma_k$
are  additional quantum numbers to define each of the ASFs and CSFs uniquely.
The  CSFs are linear combinations of Slater determinants and ASFs are 
eigenfunctions of the Dirac-Coulomb Hamiltonian $H^{\rm DC}$
\cite{grant-23-87}. In the  present study, a series of  extended optimized
level(EOL)-MCDF calculations are carried out using  GRASP92
\cite{parpia-249-96} to generate a single electron basis set  consisting of 
$(1-6)s_{1/2}$, $(2-6)p_{1/2}$, $(2-6)p_{3/2}$, $(3-5)d_{3/2}$, $4f_{5/2}$ 
and $4f_{7/2}$ orbitals. The EOL-MCDF orbitals are less state specific 
compared to ground state extremization and suitable for studying spectral
properties and structure of ASFs. Then a CSF space of a specific $J$ having
single and double excitations from a reference configuration $4f^l5d^m 6s^2$ 
to $5d$, $6p$ and $4f$ shells, where $l$ and $m$ are the occupancy of the 
shells is generated. This sequence of calculations is repeated for each of
the  atoms. Details of the $J^P$ and the reference configuration considered
and the number of CSFs  generated for each of them are given in  Table
\ref{table1}. Note that the parity ($P$) is chosen to be same as that of
the the ground state and $J$ is $4$ for even and $9/2$ for odd cases. ASFs
and corresponding eigenvalues within the CSF space considered are obtained by
a configuration interaction calculation. For further analysis related to the
structure of the ASFs, we choose CSFs which have close to uniform separation
of  $H^{\rm DC}_{kk} = \langle\gamma_kPJM|H^{\rm DC} |\gamma_k PJM\rangle$. 
For example, in Sm only 6500 of the 7325 CSFs generated are considered, the 
first 200 and last 625 CSFs are excluded. Then, the strength function 
$F_k(E) = \sum_{E'}|c_k^{E'}|^2\delta(E - E')$ of the selected CSFs are 
calculated, where $c_k^E$ is the mixing coefficient of $|\gamma_k
PJM\rangle$  for the ASF having eigenvalue $E$. To get a representative 
$\overline{F}_k(E)$ we calculate the average of $F_k(E)$s around the centroid
and over 3\% of the range of $H^{\rm DC}_{kk}$, i.e. average strength
functions with $E_k=0$.

\begin{table}
  \caption{Details of the angular momentum, the reference configuration 
considered and the number of  CSFs generated for each of the atoms. The 
numbers within parenthesis are the number of CSFs chosen for the
final calculations.}
  \label{table1}
  \begin{ruledtabular}
  \begin{tabular}{ccccc}
    Element & $J^P$       & Ref Config     & Number of CSFs  \\ 
\hline
      Ce    & $4^-$   & $4f5d 6s^2   $ &  373 (308)     \\
      Pr    & $9/2^-$ & $4f^36s^2    $ &  1378 (1278)   \\
      Nd    & $4^+$   & $4f^46s^2    $ &  2200 (2000)   \\
      Pm    & $9/2^-$ & $4f^56s^2    $ &  4378 (4178)   \\
      Sm    & $4^+$   & $4f^66s^2    $ &  7325 (6500)   
  \end{tabular}
  \end{ruledtabular}
\end{table}

Calculated average strength functions (with $E_k=0$) are compared with the
best fit $F_{k:BW-\cg}(E)$ in Fig. 6. In the fits, Eq. (5) is used to
eliminate $\beta$. The $\alpha$ values for each atom are given in the figure.
Firstly it is seen that $F_{k:BW-\cg}(E)$ gives excellent description of the
calculated strength functions and the BW to Gaussian transition is clearly
seen in Fig. 6 with $\alpha$ changing from 1.85 to 14 as we go from CeI to
SmI. The calculated $\gamma_2$ values are also consistent with this
transition as they change from 6.44 to 0.46. Comparing with Fig. 1, CeI and
PrI atoms are close to $\lambda \sim 0.1-0.15$ cases ($F_k$ is close to BW), 
NdI and PmI are close to $\lambda \sim 0.2-0.25$ cases ($F_k$ is intermediate
to BW and Gaussian)  and SmI is close to $\lambda \sim 0.3$ case ($F_k$ is
close to Gaussian) of the $m=6,N=12$ EGOE example. 
For further confirming the BW to Gaussian transition the
$\xi_2(E=0)/\xi_2^{GOE}$ values are calculated using Eq. (14) and the deduced
$\alpha$ values (used also are the calculated $\zeta$ values). They change
from 0.21 to 0.6 for CeI to SmI. These EGOE(1+2) numbers are close to
$\xi_2(E=0)/\xi_2^{GOE}$'s generated by the calculated atomic eigenstates.
However there are large fluctuations in $\xi_2(E)$, as the atomic
calculations produce in general, for many states, more localization than
expected from  EGOE(1+2); if we average $\xi_2(E)$ in the neighborhood of
$E=0$, then the calculated values are $\sim 20-30\%$ smaller than the values
given by Eq. (14). This is already seen in CeI and PrI in 
\cite{cummings-3407-01} and SmI in \cite{An-03}. The
present calculations confirm this to be a generic behavior.  The source of
this localization and modifications of EGOE(1+2) for incorporating this
property will be discussed elsewhere. Here it suffices to conclude that, from
the results in Fig. 6, CeI to SmI unmistakingly exhibit BW to Gaussian
transition for $E_k=0$.

\section{Duality between weak and strong mixing limits}

For each realization of the Hamiltonian  $H=h(1)+\lambda V(2)$, two
asymptotic natural basis emerge:  the noninteracting basis defined by $h$,
and the $\lambda=\infty$ basis defined by $V$. The previous discussions in
Sections 2 and 3 were concerned with strength functions, PR and $S^{\rm
info}$ in the noninteracting basis only. Here we extend this discussion to
the $\lambda=\infty$ basis, and will focus on the existence of a duality 
transformation between the two basis, following the recent work of Jacquod
and Varga (JV) \cite{Ja-02}. JV found that a {\it duality} point
$\lambda_d$ exists where all the statistical wavefunction properties in
these two basis coincide, and that the wavefunction  properties
in the noninteracting ($\lambda=0$) basis are related to those in the
$\lambda=\infty$ (fully interacting) basis by a duality transformation $\lambda
\leftrightarrow \lambda_d^2/\lambda$.  An ambiguity in JV results lied with
the fact that the existence and scaling of the duality point $\lambda_d$ were
derived within the BW approximation, while $\lambda_d$ explicitly lies
outside the BW regime. We therefore extend those theoretical arguments by
similar ones in the Gaussian approximation, but first recall JV results.

In the noninteracting basis, and in the BW regime, the strength functions
will have Lorentzian shape. To estimate its width via the golden rule,  one
first has to realize that there are,  beside the one-body spacing $\Delta$,
two important  energy scales \cite{Ab-90} :  the mean spacing between states
directly  coupled by the two-body interaction $\Delta_c^{(0)} = B_2^{(0)}/K 
\approx 4\Delta/N m^2$ and the $m$-body spacing  $\Delta_m^{(0)} =
B_m^{(0)}/d$, where $B_m^{(0)} \approx \sqrt{m} N \Delta$  is the $m$-body
band (note that $B_m^{(0)} \sim \sigma_h(m)$ with $\sigma_h(m)$ given by
(12,13)).  Note that this estimate slightly differs from that of JV where the
$m$-body bandwidth $B_m^{(0)}$ was approximated by the extremal possible
energy values, instead of the rms of the density of states. Then, the width
of the BW strength function is approximated via the golden rule as
$\Gamma^{(0)} \propto \lambda^2/\Delta_c^{(0)} \sim  \lambda^2 N m^2/\Delta$.
Finally, in the dilute limit $m \ll N$,  the PR is obtained as $\xi_2^{(0)} =
\Gamma^{(0)} / \Delta_m^{(0)} \propto  \lambda^2 m^{3/2} d / \Delta^2$.  Note
that, this result differs from the JV estimate given in \cite{Ja-02} by the
factor $m^{3/2}$ instead of $m$. As is the case for $\Gamma^{(0)}$, the
Golden rule gives a good estimate of the width $\Gamma^{(\infty)}$ of the
strength functions expressed in the $\lambda=\infty$ basis, in the BW 
regime. Following JV, it is seen that $ \Gamma^{(\infty)} \sim m (N-m)
\Delta^2/\lambda$, and one gets the PR in the $\lambda=\infty$ basis as
$\xi_2^{(\infty)} = \Gamma^{(\infty)} / \Delta_m^{(\infty)} \propto  (\Delta
/ \lambda)^2 d$. The duality point is then defined by
$\xi_2^{(0)}(\lambda_d)= \xi_2^{(\infty)}(\lambda_d)$ which gives the
parametric dependence
\be\label{lambdad}
\lambda_d \propto \Delta/m^{3/8}.
\ee
This result \cite{Ph-03} is in better agreement with the numerical data 
presented by JV, where it was found that $\lambda_d \sim 1/m^\nu$ with $\nu
\in [0.3,0.5]$ (one has to keep in mind however, that most data were not in
the dilute limit and that $\nu$ was extracted from a restricted range of
variation of $m$); compared to the previous JV estimate $\lambda_d \propto
\Delta/m^{1/4}$, Eq.(\ref{lambdad}) is thus in better
agreement with numerical data. 

A very close estimate for $\lambda_d$ can be derived
from PR and $S^{\rm info}$ in the Gaussian domain. Just as $\zeta=\zeta_0$
in Eqs. (10) describes $S^{\rm info}$ and PR in the Gaussian domain, 
\be
\zeta_0(\lambda) = \sigma_h/\sqrt{\sigma_h^2 + \lambda^2 \sigma_V^2} =
\sqrt{(f^2 \Delta^2)/(f^2 \Delta^2 + g^2 \lambda^2)}\;\;,
\ee
it is to be expected (by extending in a straight forward manner the
arguments  in \cite{Ks-01} where $h(1)$ basis is considered) that in the
$V(2)$ basis also the PR and $S^{\rm info}$ will be given by (10) but with
$\zeta=\zeta_\infty$ where,
\be
\zeta_\infty(\lambda) = \lambda \sigma_V/\sqrt{\sigma_h^2 + \lambda^2 
\sigma_V^2} = \sqrt{(g^2 \lambda^2)/(f^2 \Delta^2 +  g^2 \lambda^2)}
\ee 
The factors $f^2$ and $g^2$ in Eqs. (18,19) are defined in (12). In Fig. 7 it
is verified that Eq. (10) with $\zeta=\zeta_\infty$ indeed describes the
numerical EGOE(1+2) results. Having demonstrated this, it is easily seen
that  the obvious condition for $S^{\rm info}$ and PR (also strength functions)
to be same in both $h(1)$ and $V(2)$ basis is
\be
\zeta_0(\lambda_d)=\zeta_\infty(\lambda_d) \Longrightarrow 
\lambda_d=|\Delta f/g|\;,\;\;\; \zeta^2(\lambda_d)=0.5
\ee
Using Fig. 3 and the condition $\zeta^2(\lambda_d)=0.5$ gives for the
$m=6,N=12$ example, $\lambda_d = 0.29$. In the dilute limit, the $m$
dependence of $\lambda_d$ follows from Eqs. (12,13,20), 
\be
\lambda_d \sim \Delta/(3m)^{1/2}
\ee
Thus the Gaussian domain arguments give $\nu$ (in $\lambda_d \sim 1/m^\nu$)
to be 0.5 unlike the improved BW domain arguments (Eq. (17)) giving 0.375. 
With $\lambda_d$ defined, a much more significant result  that follows from
(18-20) is $\zeta_\infty(\lambda)=\zeta_0(\lambda_d^2 /  \lambda)$ and thus
there is a duality in EGOE(1+2), i.e. the results in $h(1)$ and  $V(2)$ basis
are related to each other by the duality transformation   $\lambda
\rightarrow \lambda_d^2/\lambda$. As stated before, the same transformation
is also derived by JV but using  $\Gamma_k$ and $\xi_2$ in the BW domain and
this points-out its general validity. Strictly speaking $\lambda_d$ does not
lie in the BW domain nor deep into the Gaussian domain. The duality
transformation is well tested in Fig. 8 for  $F_k(E)$ and in Fig. 9 for
$S^{\rm info}(E)$. In these calculations $\lambda_d = 0.29$.  It should be
recognized that for the strength functions in Fig. 8, the variances are
$\zeta^2_0$ and $\zeta^2_\infty$ in the $h(1)$ and $V(2)$ basis respectively.
In the case of $S^{\rm info}(E)$ one sees (from Fig. 9) departures, for
$F_k(E)$ close to Gaussian, in the region well away from the centroid of $E$
and this could be because  the tails of $F_k(E)$ display exponential
localization \cite{Fl-97,Fr-96}. These disagreements are not seen in
\cite{Ja-02} as in this work only $S^{\rm info}(E=0)$ and $\xi_2(E=0)$ are
studied. It is useful to point-out that there appears to be a close
relationship between $\lambda_d$ and thermodynamics of finite quantum
systems. Using the Gaussian domain formulas (see \cite{Ks-02})  for the
thermodynamic,  information and single particle entropies, it is easily
verified that at and around $\lambda_d$, all the three entropies will be very
close to each other; numerical verification of this result is given in
\cite{Ks-02}. Therefore it is possible to define the region around
$\lambda_d$ as `thermodynamic region' for interacting particle systems as
here different definitions of thermodynamic quantities like entropy will give
same results; see \cite{Ks-02,Ho-95}.

\section{conclusions}

In this paper an attempt is made to bring completion to the analytical (in BW
and Gaussian domains) and numerical investigations, initiated by a number of
research groups, of EGOE(1+2) random matrix model for finite interacting
quantum systems. Towards this end, a function describing the BW to Gaussian
transition in strength functions is identified (Eq. (4)) and it is used to
study participation ratio and  information and structural entropy as a
function of the interaction strength. Also it is shown, using Gaussian domain
results, that the duality point $\lambda_d$  behaves more like $\lambda_d
\sim 1/\sqrt{m}$ where $m$ is number of fermions. Applications of these
results are given for the BW to Gaussian transition in the series of neutral
atoms CeI, PrI, NdI, PmI and SmI. As for EGOE(1+2), what remains is a 
rigorous analytical treatment of this random  matrix model. This will give
for example a theory for $\alpha$ vs $\lambda$ (see Fig. 2), a theory for
$S^{\rm info}$ and PR in the $\lambda \la \lambda_c$ domain etc. Finally it
is important to remind that only recently rigorous analytical treatment  has
started becoming available for the simpler EGOE(2) \cite{Be-01}.

\acknowledgments

Thanks are due to Ph. Jacquod for a careful reading of the first draft of
the paper and for making many suggestions for improving it. The present work
was initiated as a result of the correspondence one of the authors (VKBK)
have  had with Ph. Jacquod.    Thanks are also due to Imre Varga for
correspondence in the initial stages of this work.

\newpage

\begin{figure}
\includegraphics[width=8cm]{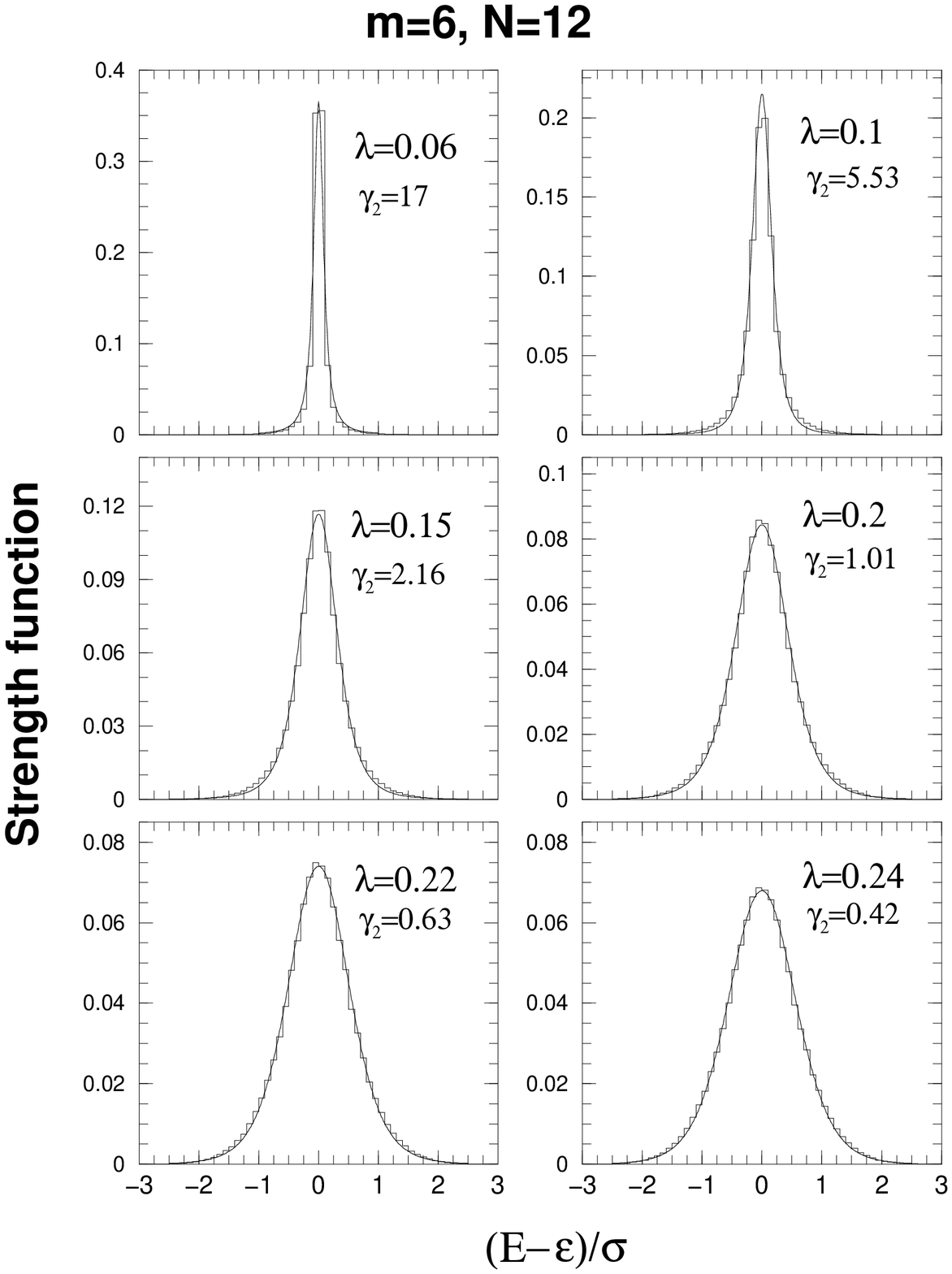}
\includegraphics[width=8cm]{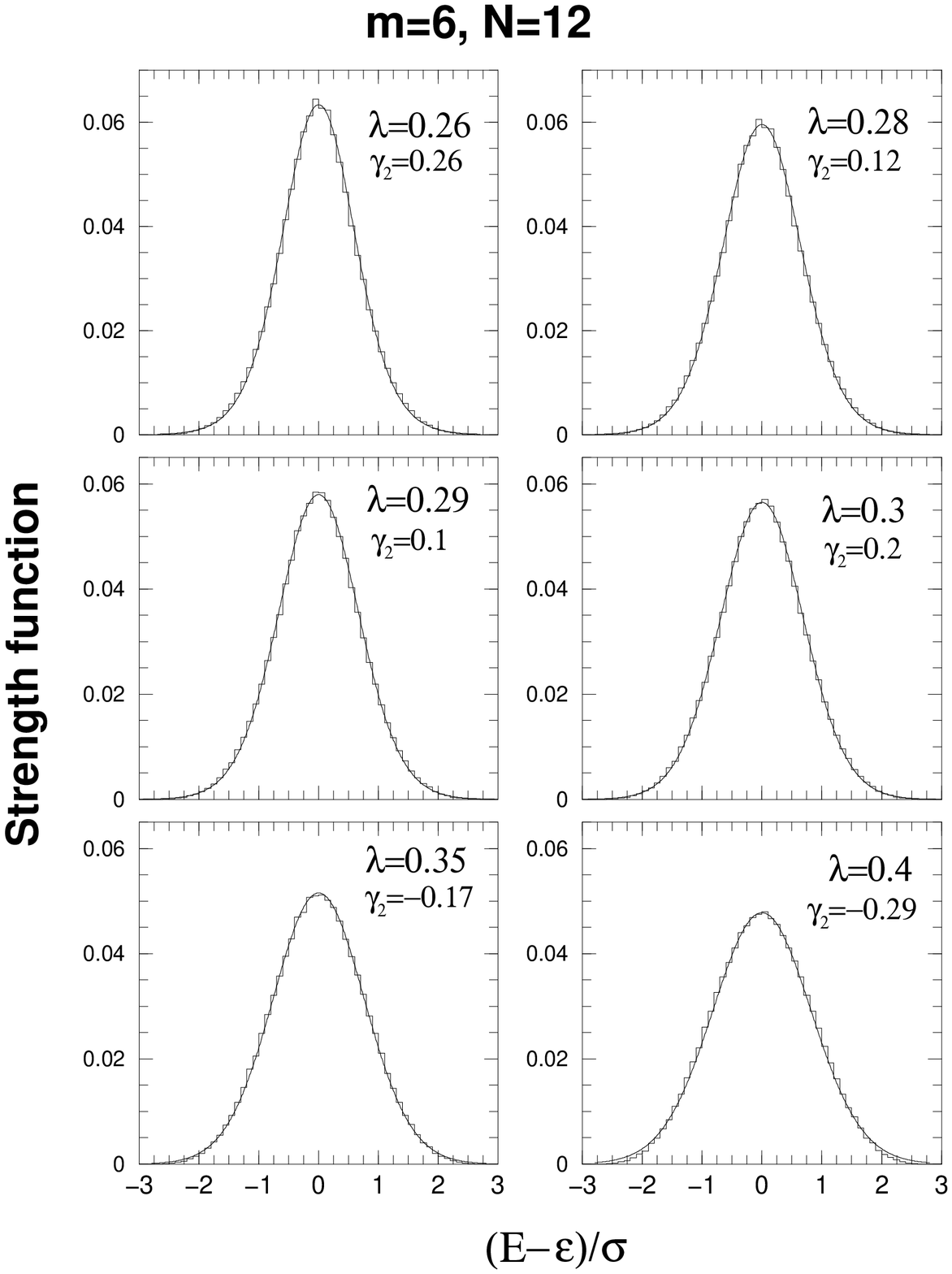}
\caption{Strength functions $F_k(E)$  for a 20 member EGOE(1+2)
for various values of the interaction strength $\lambda$ in $\{H\} = h(1) +
\lambda \{V(2)\}$ for a system of 6 fermions in 12 single particle states;
the  matrix dimension is 924. The single particle energies used in the
calculations are $\epsilon_i=(i+1/i), i=1,2,\ldots,12$ just as in
\cite{Ko-01}. In the figures $F_k(E)$ is plotted against
${\hat{E}}=(E-\epsilon)/\sigma$ where $\epsilon$ is the spectrum centroid and
$\sigma$ is the width. The histograms are EGOE(1+2) results and the
continuous curves are the best fit $F_{k:BW-\cg}(E)$ from Eq. (4). In
constructing the strength functions, $|C_k^E |^2$ are summed over the basis
states $\l.\l| k \ri. \ran$ in the energy window ${\hat{E}}_k \pm \Delta$ and
then the ensemble averaged $F_{{\hat{E}}_k} ({\hat{E}})$ vs ${\hat{E}}$ is
constructed as a histogram; the value of $\Delta$ is chosen to be 0.05 for
$\lambda = 0.06$ and beyond this $\Delta=0.1$. Here ${\hat{E}}_k =
(E_k-\epsilon_H)/\sigma_H$ and in the figures results shown for $F_k(E)$ with
${\hat{E}}_k=0$. Similar results are also obtained for the $(m=7,N=14)$
system.}
\end{figure}


\begin{figure}
\includegraphics[width=14cm]{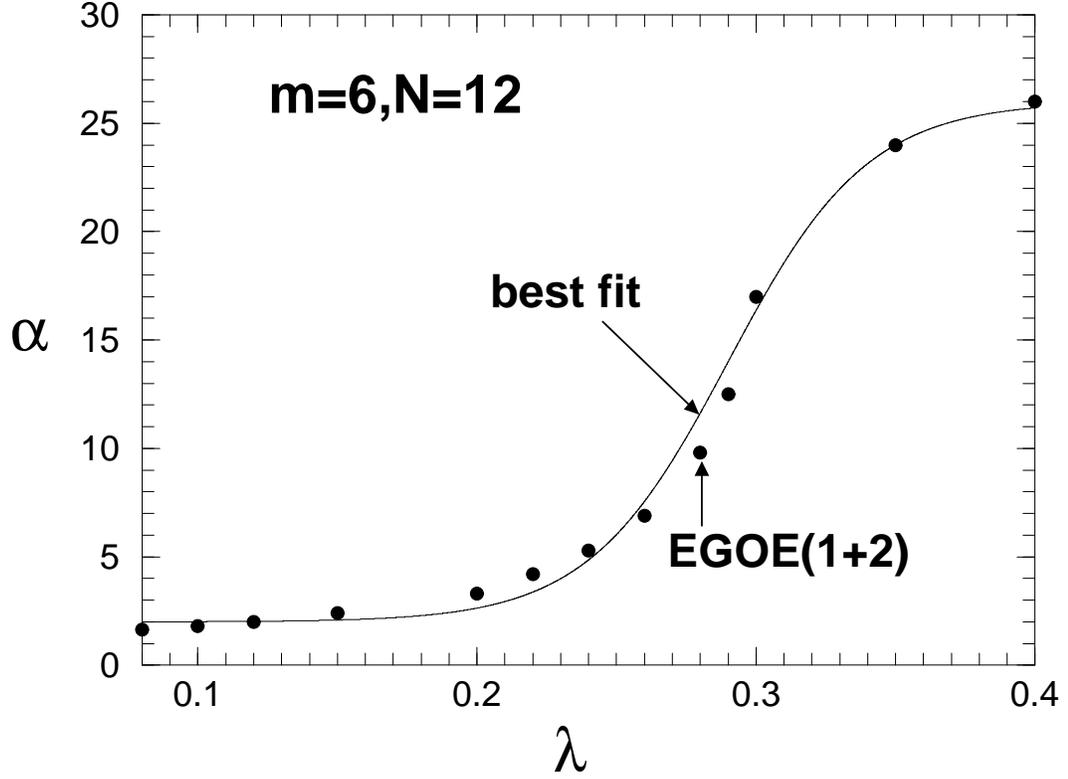}
\caption{For $\lambda \geq 0.08$,
$\alpha$ vs $\lambda$ obtained by fitting the $F_k(E)$ in $h(1)$ basis to the
interpolating form $F_{k:BW-\cg}$ given by (4). Results are shown for the
EGOE(1+2) system used in Fig. 1.  The filled circles give the best fit
$\alpha$ values and the continuous curve, given by
$\alpha=24/[exp-(40(\lambda-\lambda_0))+1]+2$ with $\lambda_0=0.29$, 
guides the eye. It is curious to note that $\lambda_0$ is close to
$\lambda_d$, the duality pont discussed in Section 5.
Similar results are also obtained for the $(m=7,N=14)$ system.}
\end{figure}


\begin{figure}
\includegraphics[width=14cm]{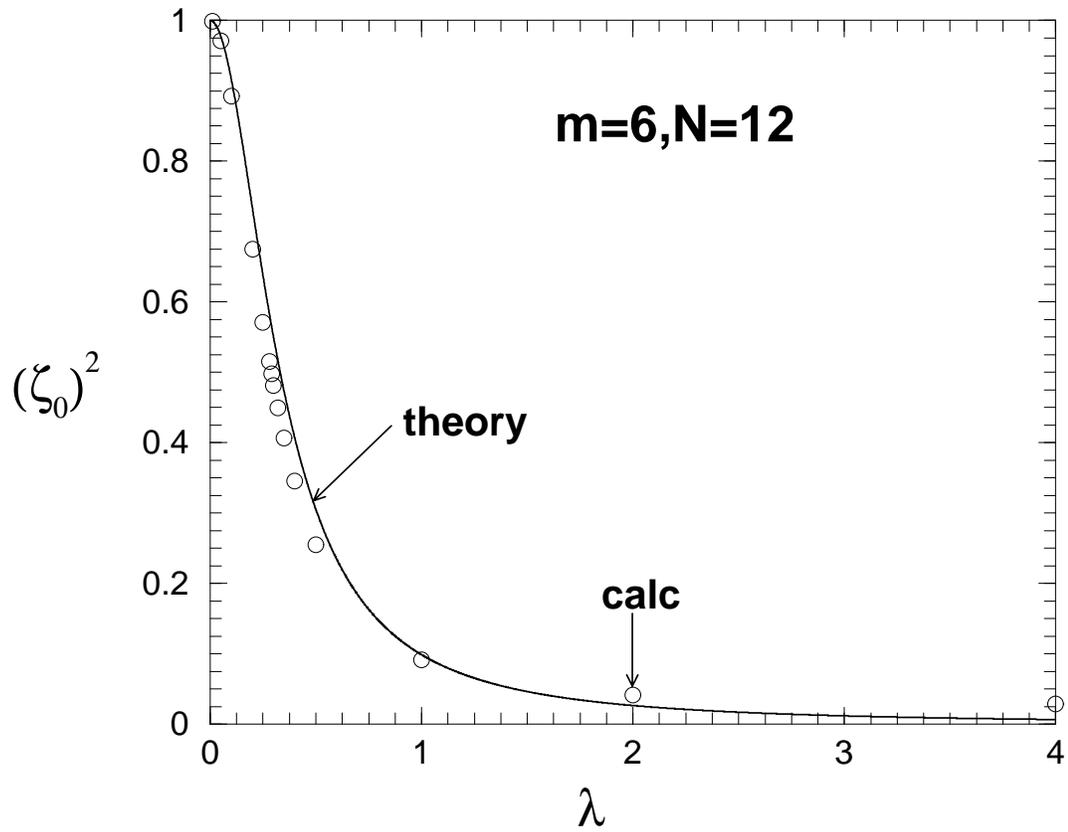}
\caption{Square of the correlation coefficient $\zeta_0^2$ vs 
$\lambda$ for the EGOE(1+2) system used in Fig. 1. Theoretical results
(continuous curve) given by (11,12) are compared with the numerical EGOE(1+2)
results  (open circles).}
\end{figure}


\begin{figure}
\includegraphics[width=14cm]{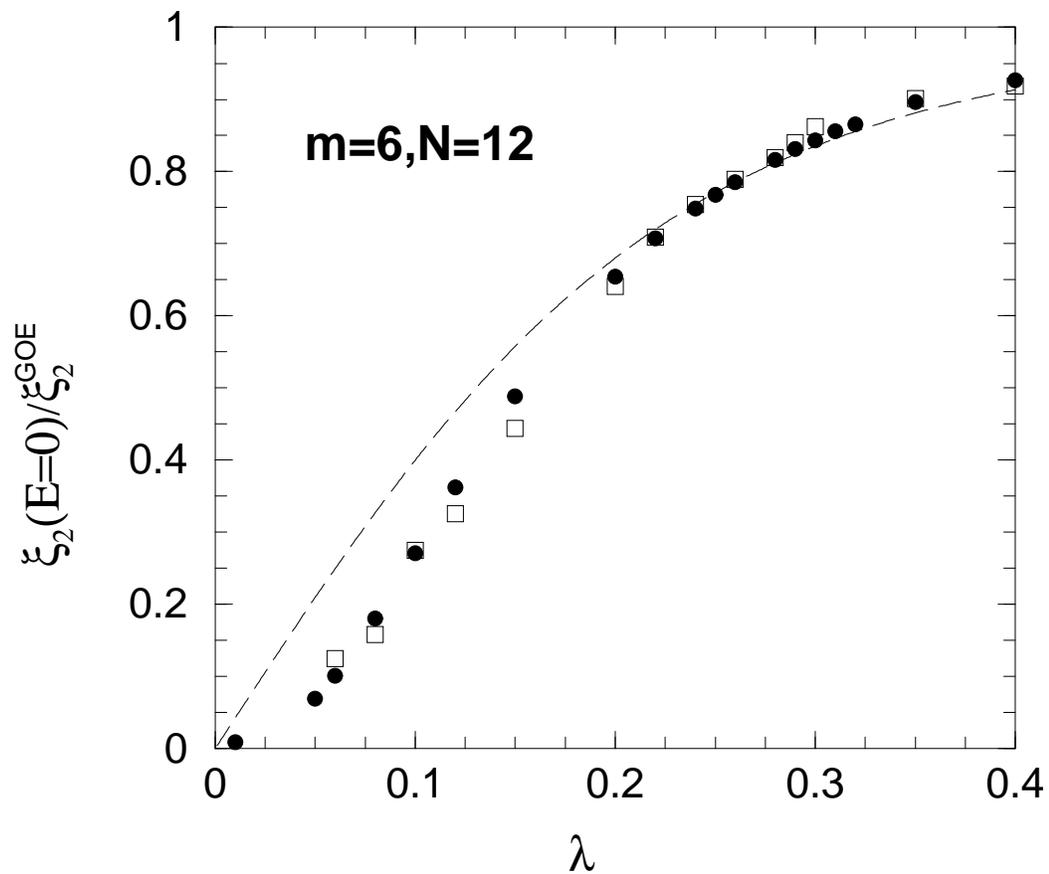}
\caption{Participation ratio $\xi_2(E=0)/\xi_2^{\rm GOE}$ vs
$\lambda$ for the  EGOE(1+2) system used in Fig. 1. Theoretical results given
by (14) (open squares) are compared with the EGOE(1+2) results (filled
circles). For comparison, the Gaussian domain result (dashed curve) from (10)
is also shown.}
\end{figure}


\begin{figure}
\includegraphics[width=14cm]{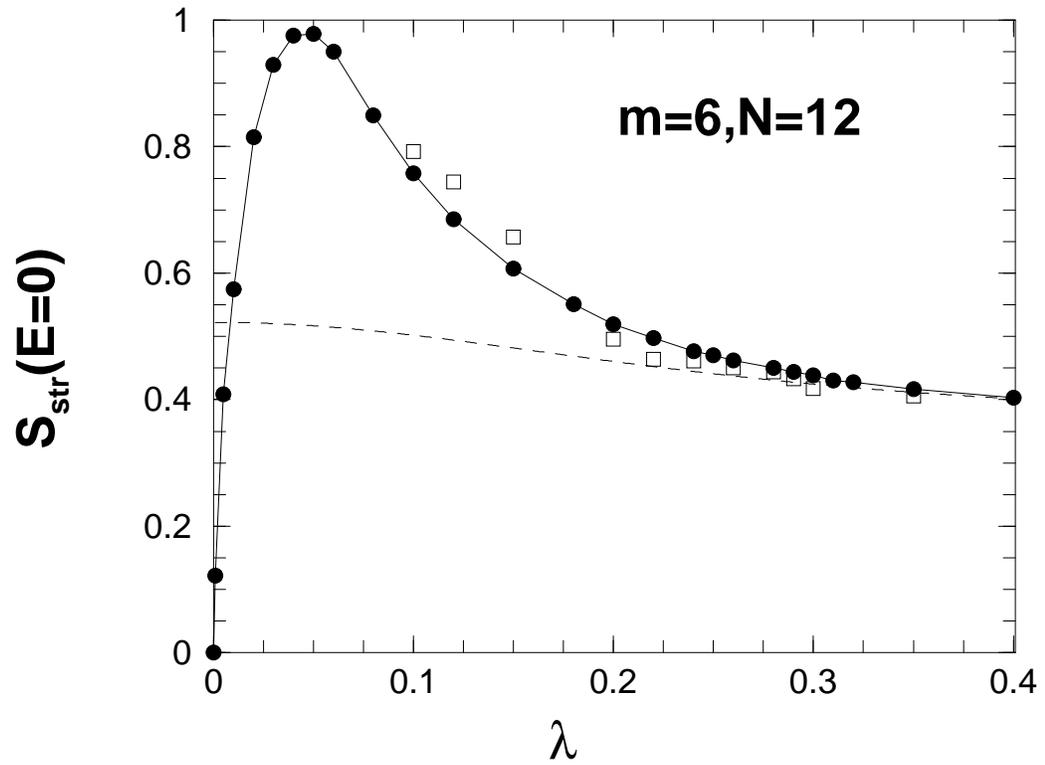}
\caption{Structural entropy $S_{\rm str}(E=0)$ vs $\lambda$ for
the   EGOE(1+2) system used in Fig. 1.  Theoretical results given by (14,8)
(open squares) are compared with the EGOE(1+2) results (filled circles). For
comparison, the Gaussian domain result (dashed curve) from (16) is also
shown.  }
\end{figure}


\begin{figure}
\includegraphics[width=10cm]{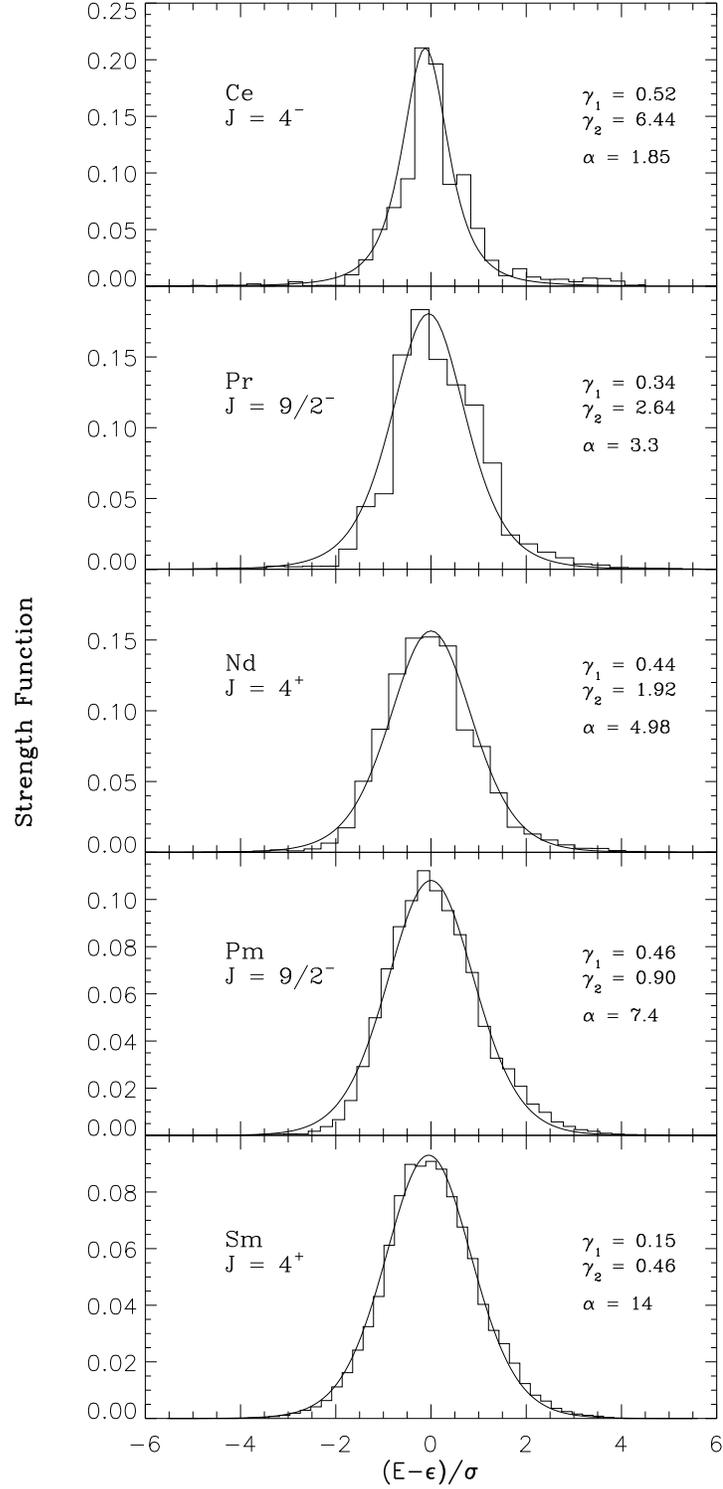}
\caption{Strength functions $F_k(E)$ for CeI to SmI. Histograms are 
calculated
strength functions and the smooth curves are $F_{k:BW-\cg}(E)$. Also given in
the figure are the calculated $\gamma_1$ (skewness) and $\gamma_2$ (excess) 
values and the deduced values, from the best fits, of $\alpha$ characterizing
$F_{k:BW-\cg}(E)$.}
\end{figure}


\begin{figure}
\includegraphics[width=14cm]{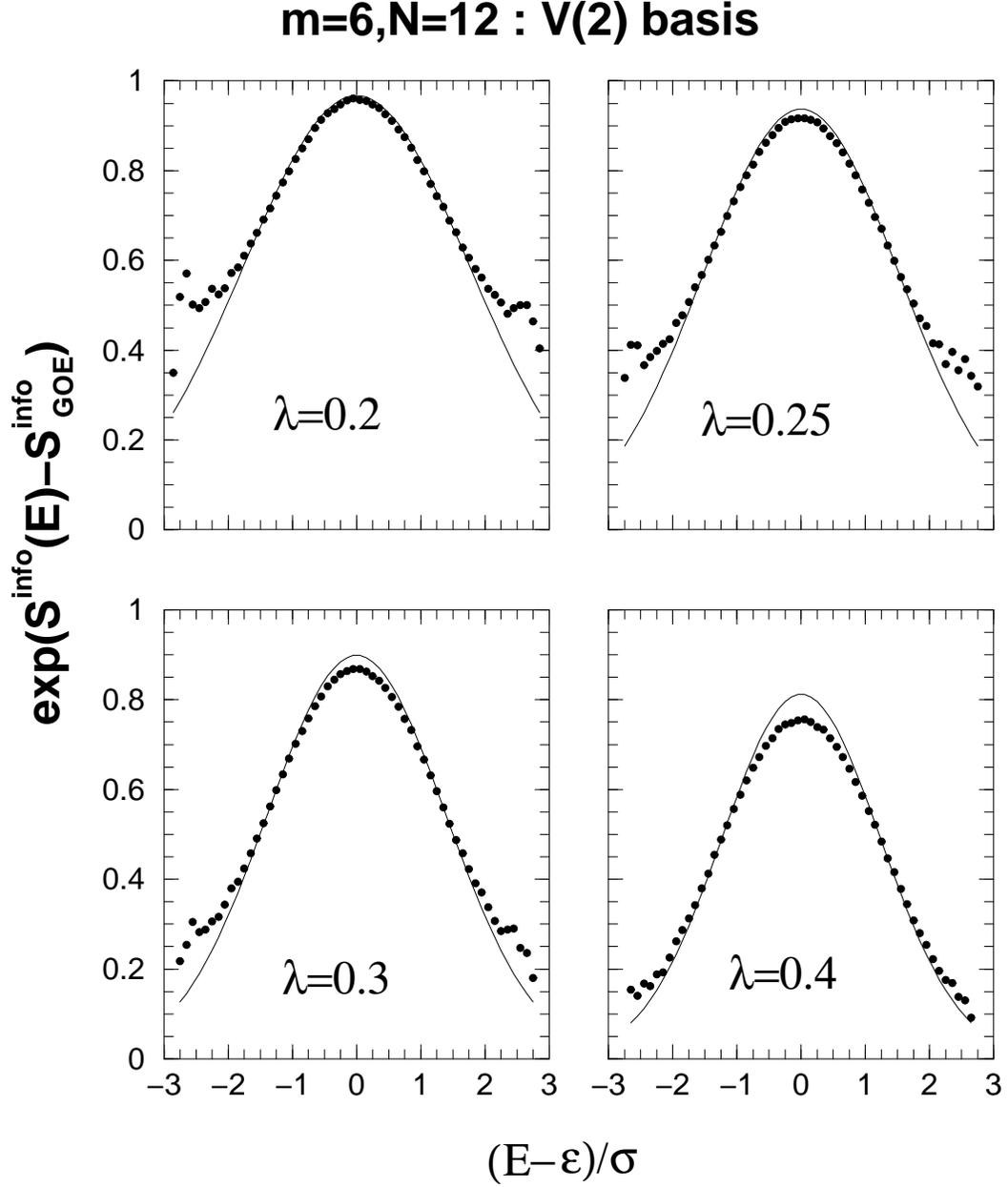}
\caption{$\exp(S^{\rm info}(E=0) - S^{\rm info}_{GOE})$ in V(2) basis
(filled circles) for four different $\lambda$ values for the EGOE(1+2) system
used in Fig. 1. The continuous curve is from Eq. (10) with 
$\zeta=\zeta_\infty$.}
\end{figure}


\begin{figure}
\includegraphics[width=14cm]{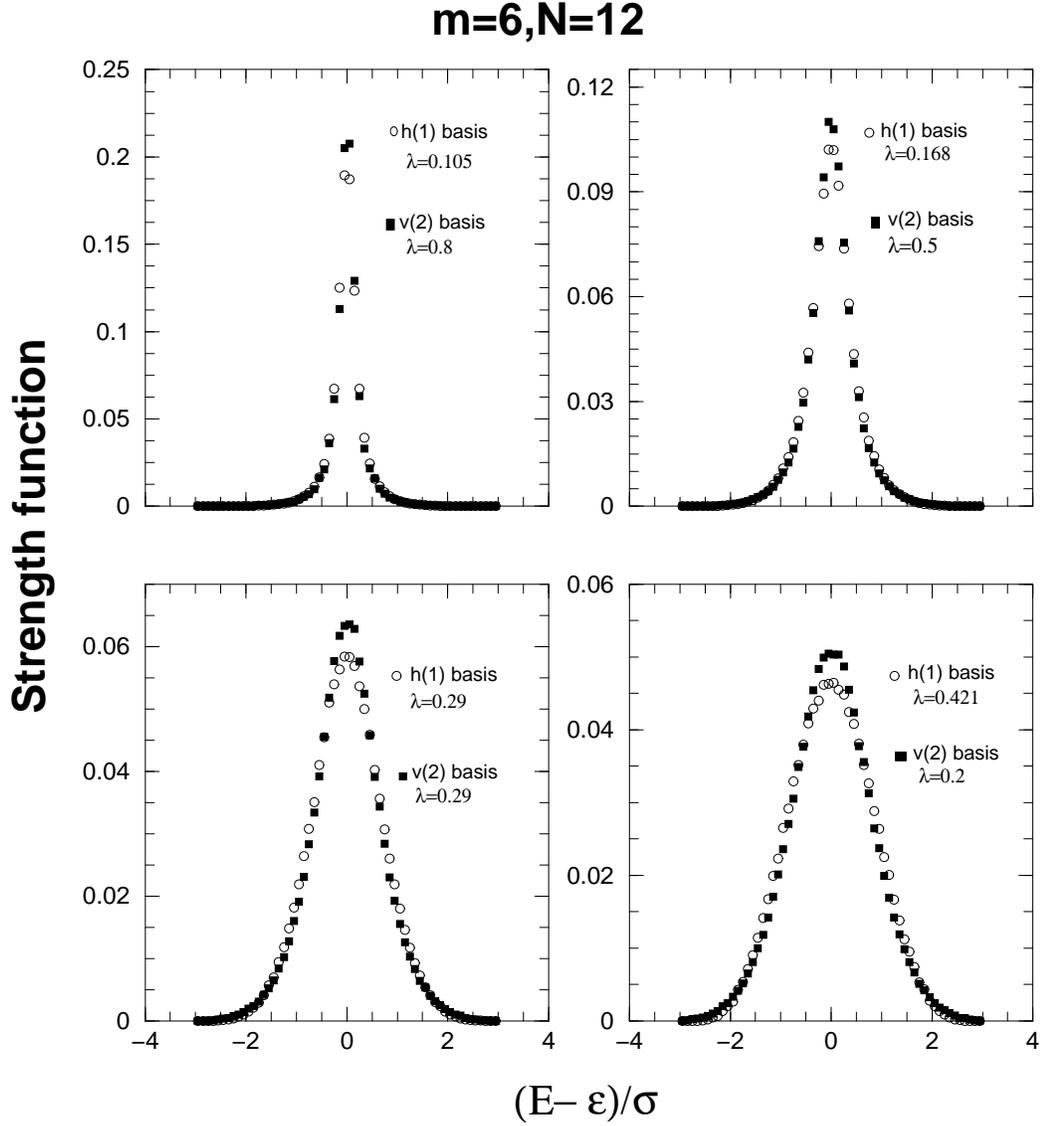}
\caption{Strength functions $F_k(E)$ in the $h(1)$ and $V(2)$
basis for four $\lambda$ values related by the duality transformation 
$\lambda \rightarrow \lambda_d^2 / \lambda$. Results are for the EGOE(1+2) 
system used in Fig. 1. Here $\lambda_d=0.29$. Similar results for the BW
spreading widths  are given in \cite{Ja-02} for several EGOE(1+2) systems
with $h(1)$ also chosen to be random.}
\end{figure}


\begin{figure}
\includegraphics[width=14cm]{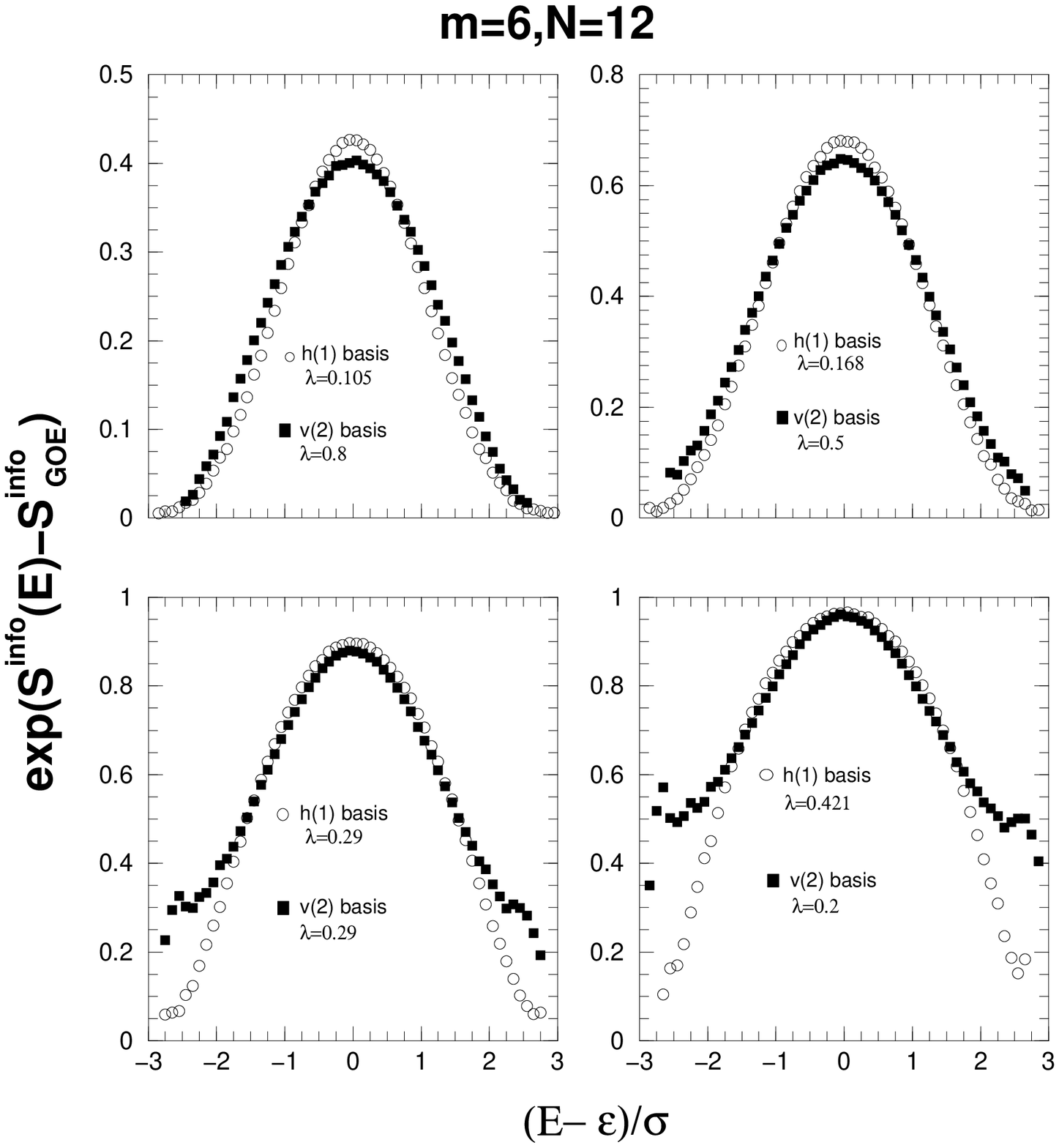}
\caption{ Same as Fig. 7 but for $\exp(S^{\rm info}(E) - S^{\rm info}_{
GOE})$. Similar results for $\xi_2$ but only at $E=0$ are given   in
\cite{Ja-02}.}
\end{figure}

\ed
\begin{thebibliography}{99}

\bibitem{Ko-01} V.K.B. Kota, Phys. Rep. {\bf 347}, 223 (2001).

\bibitem{Bro-81} T.A. Brody, J. Flores, J.B. French, P.A. Mello, 
A. Pandey, and S.S.M. Wong, Rev. Mod. Phys. {\bf 53}, 385-479 (1981).

\bibitem{Ks-01} V.K.B. Kota and R. Sahu, Phys. Rev. E {\bf 64}, 016219 (2001).

\bibitem{Ze-02} J.M.G. G\'omez, K. Kar, V.K.B. Kota, J. Retamosa, and R.
Sahu, Phys. Rev.  C {\bf 64}, 034305 (2001); V. Vel\'azquez  and A.P. Zuker,
Phys. Rev. Lett. {\bf 88}, 072502 (2002); M. Horoi, J. Kaiser, and V.
Zelevinsky, Phys. Rev. C {\bf 67}, 054309 (2003); V.K.B. Kota, Ann. Phys.
(N.Y.) {\bf 306}, 58 (2003).

\bibitem{Fl-99} V.V. Flambaum, A.A. Gribakina, G.F. Gribakin, and I.V.
Ponomarev, Physica D {\bf 131}, 205 (1999); V.V. Flambaum, A.A.
Gribakina, G.F. Gribakin, and C. Harabati, Phys. Rev. A {\bf 66}, 012713
(2002).

\bibitem{An-03} Dilip Angom and V.K.B. Kota, Phys. Rev. A {\bf 67}, 052508 
(2003).

\bibitem{Ja-01}  X. Leyronas, P.G. Silvestrov, and C.W.J. Beenakker, Phys.
Rev. Lett. {\bf 84}, 3414 (2000); Ph. Jacquod and A.D. Stone, Phys. Rev.
Lett. {\bf 84}, 3938  (2000); Phys. Rev. B {\bf 64}, 214416 (2001).

\bibitem{Al-00}  Y. Alhassid, Ph. Jacquod, and A. Wobst, Phys. Rev. B {\bf
61}, R13357 (2000), Physica {\bf E9}, 393 (2001); Y. Alhassid and A. Wobst,
Phys. Rev. B {\bf 65}, 041304 (2002).

\bibitem{Pa-02}   T. Papenbrock, L. Kaplan, and G.F. Bertsch, Phys. Rev. B
{\bf 65}, 235120 (2002).

\bibitem{Mez-87} M. M\'ezard, G. Parisi, and M.A. Virasoro, Spin
Glass Theory and Beyond (World Scientific, Singapore, 1987).

\bibitem{Qc-00} B. Georgeot and D.L. Shepelyansky, Phys. Rev. E {\bf
62}, 3504 (2000); {\bf 62}, 6366 (2000);  G. Benenti, G. Casati, and 
D.L. Shepelyansky, Euro. Phys. J. {\bf D17}, 265 (2001); V.V. Flambaum and
F.M. Izrailev, {\it ibid.} {\bf 64}, 026124 (2001).

\bibitem{Kok-02} V.K.B. Kota and K. Kar, Phys. Rev. E {\bf 65}, 026130
(2002).

\bibitem{Fl-96} V.V. Flambaum, G.F. Gribakin, and F.M. Izrailev, Phys.
Rev. E {\bf 53}, 5729 (1996). 

\bibitem{Fl-97} V.V. Flambaum and F.M. Izrailev, Phys. Rev. E  {\bf 56}, 
5144 (1997).

\bibitem{Ab-90} S. \r{A}berg, Phys. Rev. Lett. {\bf 64}, 3119 (1990).

\bibitem{Ja-97} Ph. Jacquod and D.L. Shepelyansky, Phys. Rev. Lett. 
{\bf 79}, 1837 (1997).

\bibitem{Gs-97} B. Georgeot and D.L. Shepelyansky, Phys. Rev. Lett. {\bf 79}, 
4365 (1997).

\bibitem{Fl-00} V.V. Flambaum and F.M. Izrailev, Phys. Rev. E  {\bf 61},
2539 (2000); V.K.B. Kota and R. Sahu, preprint nucl-th/0006079.

\bibitem{Ja-02} Ph. Jacquod and I. Varga, Phys. Rev. Lett. {\bf 89},  134101
(2002).

\bibitem{Ks-02} V.K.B. Kota and R. Sahu, Phys. Rev. E {\bf 66}, 037103
(2002).

\bibitem{Mehta} M.L. Mehta, Random Matrices, 2nd edition (Academic 
Press, New York, 1991).

\bibitem{Ja-95} Ph. Jacquod and D.L. Shepelyansky, Phys. Rev. Lett. {\bf
75}, 3501 (1995).

\bibitem{Be-01} L. Benet, T. Rupp, and H.A. Weidenm\"{u}ller, Phys. Rev.
Lett. {\bf 87}, 010601 (2001); Ann. Phys.  (N.Y.) {\bf 292}, 67 (2001); Z.
Pluhar and H.A. Weidenm\"{u}ller, Ann. Phys.  (N.Y.), {\bf 297}, 344 (2002).

\bibitem{Ke-69} A. Stuart and J.K. Ord, Kendall's Advanced Theory of 
Statistics, fifth edition of Volume 1: Distribution Theory (Oxford
University Press, New York, 1987). 

\bibitem{Im-02} I. Varga and J. Pipek, Phys. Rev. E {\bf 68}, 026202 (2003). 

\bibitem{Ab-64} M. Abramowtiz, I.A. Stegun (Eds.),  Handbook of
Mathematical functions, NBS Applied Mathematics Series, Vol. 55,
U.S. Govt. Printing Office, Washington, D.C. (1964).

\bibitem{cummings-3407-01} A.~Cummings, G.~O'Sullivan, and D.~M.~Heffernan,
J. Phys. \textbf{B34}, 3407 (2001).

\bibitem{angom-271-2001} Angom Dilip,  I. Endo, A. Fukumi, M. Linuma, T.
Kondo, and T. Takahasi, Euro. Phys. J.
\textbf{D14}, 271 (2001).

\bibitem{sekiya-012503-01} M. Sekiya, K. Narita, and H. Tatewaki,
Phys. Rev. A \textbf{63}, 012503 (2001).
  
\bibitem{grant-23-87} I.~P.~Grant and H.~M.~Quiney, in Advances in Atomic
and Molecular Physics, Vol 23, edited by  D.~Bates and B.~Bederson 
(Academic Press, New York, 1987) p. 37.

\bibitem{parpia-249-96} F.~Parpia, C.~Fischer, and I.~Grant,
Comput. Phys. Commun. \textbf{94}, 249 (1996).

\bibitem{Ph-03} Ph. Jacquod, private communication (2003).

\bibitem{Fr-96} N. Frazier, B.A. Brown, and V. Zelevinsky, Phys. Rev. C 
{\bf 54}, 1665 (1996); W. Wang, F.M. Izrailev, G. Casati, Phys. Rev. E 57
(1998) 323.

\bibitem{Ho-95} M. Horoi, V. Zelevinsky, and B.A. Brown, Phys. Rev. Lett.
{\bf 74},  5194 (1995).

\end{thebibliography}
